\documentstyle[twoside]{article}

\catcode`\@=11
\long\def\@makefntext#1{
\protect\noindent \hbox to 3.2pt {\hskip-.9pt  
$^{{\eightrm\@thefnmark}}$\hfil}#1\hfill}		

\def\@makefnmark{\hbox to 0pt{$^{\@thefnmark}$\hss}}	
	
\def\ps@myheadings{\let\@mkboth\@gobbletwo
\def\@oddhead{\hbox{}
\rightmark\hfil\eightrm\thepage}   
\def\@oddfoot{}\def\@evenhead{\eightrm\thepage\hfil
\leftmark\hbox{}}\def\@evenfoot{}
\def\sectionmark##1{}\def\subsectionmark##1{}}

\oddsidemargin=\evensidemargin
\newcounter{sectionc}\newcounter{subsectionc}\newcounter{subsubsectionc}
\renewcommand{\section}[1] {\vspace{12pt}\addtocounter{sectionc}{1} 
\setcounter{subsectionc}{0}\setcounter{subsubsectionc}{0}\noindent 
	{\tenbf\thesectionc. #1}\par\vspace{5pt}}
\renewcommand{\subsection}[1] {\vspace{12pt}\addtocounter{subsectionc}{1} 
	\setcounter{subsubsectionc}{0}\noindent 
	{\bf\thesectionc.\thesubsectionc. {\kern1pt \bfit #1}}\par\vspace{5pt}}
\renewcommand{\subsubsection}[1] {\vspace{12pt}\addtocounter{subsubsectionc}{1}
	\noindent{\tenrm\thesectionc.\thesubsectionc.\thesubsubsectionc.
	{\kern1pt \tenit #1}}\par\vspace{5pt}}
\newcommand{\nonumsection}[1] {\vspace{12pt}\noindent{\tenbf #1}
	\par\vspace{5pt}}

\newcommand{\textlineskip}{\baselineskip=13pt}
\newcommand{\smalllineskip}{\baselineskip=10pt}

\def\eightcirc{
\begin{picture}(0,0)
\put(4.4,1.8){\circle{6.5}}
\end{picture}}
\def\eightcopyright{\eightcirc\kern2.7pt\hbox{\eightrm c}} 

\newcommand{\copyrightheading}[1]
        {\vspace*{-2.5cm}\smalllineskip{\flushleft
        {\footnotesize Nuovo Cimento B 114, 717-722, June 1999 #1}\\
       {\footnotesize Los Alamos electronic archives: cond-mat/9706193 #1}\\
        {\footnotesize $\eightcopyright$\,
        Nuovo Cimento, Reyes \& Rosu
        }\\
         }}

\def\abstracts#1#2#3{{
	\centering{\begin{minipage}{4.5in}\baselineskip=10pt\footnotesize
	\parindent=0pt #1\par 
	\parindent=15pt #2\par
	\parindent=15pt #3
	\end{minipage}}\par}} 

\renewenvironment{thebibliography}[1]
	{\frenchspacing
	 \ninerm\baselineskip=11pt
	 \begin{list}{\arabic{enumi}.}
        {\usecounter{enumi}\setlength{\parsep}{0pt}     
	 \setlength{\leftmargin 12.7pt}{\rightmargin 0pt} 
         \setlength{\itemsep}{0pt} \settowidth
	{\labelwidth}{#1.}\sloppy}}{\end{list}}

\newcounter{itemlistc}
\newcounter{romanlistc}
\newcounter{alphlistc}
\newcounter{arabiclistc}

\def\@citex[#1]#2{\if@filesw\immediate\write\@auxout
	{\string\citation{#2}}\fi
\def\@citea{}\@cite{\@for\@citeb:=#2\do
	{\@citea\def\@citea{,}\@ifundefined
	{b@\@citeb}{{\bf ?}\@warning
	{Citation `\@citeb' on page \thepage \space undefined}}
	{\csname b@\@citeb\endcsname}}}{#1}}

\newif\if@cghi
\def\cite{\@cghitrue\@ifnextchar [{\@tempswatrue
	\@citex}{\@tempswafalse\@citex[]}}
\def\citelow{\@cghifalse\@ifnextchar [{\@tempswatrue
	\@citex}{\@tempswafalse\@citex[]}}
\def\@cite#1#2{{$\null^{#1}$\if@tempswa\typeout
	{IJCGA warning: optional citation argument 
	ignored: `#2'} \fi}}

\def\@refcitex[#1]#2{\if@filesw\immediate\write\@auxout
	{\string\citation{#2}}\fi
\def\@citea{}\@refcite{\@for\@citeb:=#2\do
	{\@citea\def\@citea{, }\@ifundefined
	{b@\@citeb}{{\bf ?}\@warning
	{Citation `\@citeb' on page \thepage \space undefined}}
	\hbox{\csname b@\@citeb\endcsname}}}{#1}}

\def\@refcite#1#2{{#1\if@tempswa\typeout
        {IJCGA warning: optional citation argument
	ignored: `#2'} \fi}}

\def\refcite{\@ifnextchar[{\@tempswatrue
	\@refcitex}{\@tempswafalse\@refcitex[]}}

\def\pmb#1{\setbox0=\hbox{#1}
	\kern-.025em\copy0\kern-\wd0
	\kern.05em\copy0\kern-\wd0
	\kern-.025em\raise.0433em\box0}

\def\fnt#1#2{\footnotetext{\kern-.3em
	{$^{\mbox{\scriptsize #1}}$}{#2}}}

\font\tenrm=cmr10
\font\tenit=cmti10 
\font\tenbf=cmbx10
\font\bfit=cmbxti10 at 10pt
\font\ninerm=cmr9

\font\eightrm=cmr8

\def\qed{\hbox{${\vcenter{\vbox{			
   \hrule height 0.4pt\hbox{\vrule width 0.4pt height 6pt
   \kern5pt\vrule width 0.4pt}\hrule height 0.4pt}}}$}}

\begin{document}



\normalsize\textlineskip
\thispagestyle{empty}
\setcounter{page}{1}

\copyrightheading{}                     

\vspace*{0.88truein}

\centerline{\bf General solution of the three-site master equation and the
discrete Riccati equation
     }
\vspace*{0.035truein}
\vspace*{0.37truein}
\centerline{\footnotesize M. Reyes and H. Rosu}
\vspace*{0.015truein}
\centerline{\footnotesize\it Instituto de F\'{\i}sica,
Universidad de Guanajuato, Apdo Postal E-143, 37150 Le\'on, Gto, Mexico}
\vspace*{0.225truein}

\vspace*{0.21truein}
\abstracts{
{\bf Summary:} - We first obtain by analogy with the
continuous (differential) case the general solution of the
discrete Riccati equation. Moreover, we establish the
full equivalence between our discrete Riccati equation and a corresponding
homogeneous second order discrete linear equation.
Our results can be considered
the discrete analog of Mielnik's construction in supersymmetric quantum
mechanics [J. Math. Phys. {\bf 25}, 3387 (1984)].
We present an application to the three-site
master equation obtaining explicitly the general solutions for the simple cases
of free random walk and the biased random walk.
}{}{}


\textlineskip                  
\vspace*{12pt}                 

\vspace*{1pt}\textlineskip	
\vspace*{-0.5pt}
\noindent


\noindent




\noindent










\noindent
We consider the continuous Riccati equation (CRE)
$y^{'}=a(x)y^2+b(x)y+c(x)$
with the known particular solution $y_{0}$ and let $y_1=u+y_{0}$ be the
second solution.
By substituting $y_{1}$ in CRE one gets the Bernoulli equation
$u^{'}=au^2+(2ay_{0}+b)u$. Furthermore,
using $v=1/u$, the simple first-order linear differential equation
$v^{'}+(2ay_{0}+b)v+a=0$ is obtained, which can be solved by employing the
integration factor $f=e^{\int ^{x}2ay_{0}+b}$, leading
to the solution
$v=e^{-\int ^{x}2ay_{0}+b}(-\int ^{x}ae^{\int ^{x}2ay_{0}+b}+C)$,
where $C$ is an integration constant. Thus, the
general solution of CRE is
$$
y_{1}=y_{0}+
\frac{e^{\int ^{x}2ay_{0}+b}}{C-\int ^{x}ae^{\int ^{x}2ay_{0}+b}}\equiv
y_{0}+\frac{f}{C-\int ^{x}(af)}~.
\eqno(1)
$$
For the usual nonrelativistic one-dimensional quantum mechanics
$b=0$ and $a=-1$, i.e., the integration factor is
$f_{qm}=e^{-2\int ^{x}y_{0}}$.  The particular solution $y_{0}$ is known as
Witten's superpotential [\refcite{W}] in the context of supersymmetric quantum
mechanics [\refcite{SQM}], while the general solution
$y_1=y_{0}+\frac{f_{qm}}{C+\int ^{x}f_{qm}}$ may be called
Mielnik's superpotential [\refcite{M3}]. Moreover, $f_{qm}^{1/2}$ is the
ground state wavefunction, $-2y_{0}^{'}$
($\equiv -2\frac{d^2}{dx^2}\ln f_{qm}^{1/2}$)
is the Darboux transform part of the
Schr\"odinger potential and $\frac{f_{qm}^{1/2}}{C+\int^{x} f_{qm}}$ is
the ground state wavefunction corresponding to Mielnik's superpotential for
which $-2y_{1}^{'}$ can be thought of as the general Darboux transform part
in the potential.
All these relationships make up the core of the ``entanglement"
between Riccati and Schr\"odinger equations,
which has been recently emphasized by Haley [\refcite{h97}].

Consider now the discrete Riccati equation (DRE) $y_{n+1}=a_{n}y_{n+1}y_{n}
+b_{n}y_{n}+c_{n}$, with known solution $y_{n}^{0}$. As in the case of CRE,
the general
solution is chosen of the form $y_{n}^{1}=y_{n}^{0}+u_{n}$. The discrete
equation for the function $u_{n}$ (discrete Bernoulli equation) will be
$$
u_{n+1}=\frac{a_{n}}{1-a_{n}y_{n}^{0}}u_{n+1}u_{n}+
\frac{b_{n}+a_{n}y_{n+1}^{0}}{1-a_{n}y_{n}^{0}}u_{n}~.
\eqno(2)
$$
We now take $v_{n}=1/u_{n}$ to get
$$
v_{n+1}=\Bigg(\frac{1-a_{n}y_{n}^{0}}{b_{n}+a_{n}y_{n+1}^{0}}\Bigg)v_{n}
-\frac{a_{n}}{b_{n}+a_{n}y_{n+1}^{0}}=g_{n}v_{n}+h_{n}~,
\eqno(3)
$$
whose solution reads
$$
v_{n+1}=\Bigg(\prod _{k=0}^{n}g_{k}\Bigg)\Bigg(D-\sum _{k=0}^{n}
\Bigg(\prod _{j=0}^{k}g_{j}\Bigg)^{-1}h_{k}\Bigg)~,
\eqno(4)
$$
where $D$ is a positive constant.
Thus, the general DRE solution can be written down as follows
$$
y_{n}^{1}=y_{n}^{0}+
\frac{\prod _{k=0}^{n-1}
(\frac{b_{k}+a_{k}y_{k+1}^{0}}{1-a_{k}y_{k}^{0}})}
{
D-\sum _{k=0}^{n-1}\Bigg(\prod _{j=0}^{k}
\frac{b_{j}+a_{j}y_{j+1}^{0}}{1-a_{j}y_{j}^{0}}
\Bigg)\frac{a_{k}}{b_{k}+a_{k}y_{k+1}^{0}}}~.
\eqno(5a)
$$
One can remark the formal analogy between the CRE and DRE general solutions.
Indeed, $f_{p}=\prod_{l=0}^{p}\frac{b_{l}+a_{l}y_{l+1}^{0}}{1-a_{l}y_{l}^{0}}$
can be thought of as a discrete ``integration" factor (DIF), and thus
Eq.~(5a) can be written in the form
$$
y_{n}^{1}=y_{n}^{0}+\frac{f_{n-1}}{D-\sum_{k=0}^{n-1}
\frac{f_{k}a_{k}}{b_{k}+a_{k}y_{k+1}^{0}}}~,
\eqno(5b)
$$
from which the analogy (and difference) with respect to the continuous case
is transparent.

We now want to establish a formal equivalence between the DRE and
the homogeneous second-order discrete equation (HSODE), more exactly, between
$y_{n+1}=a_{n}y_{n+1}y_{n}+b_{n}y_{n}+c_{n}$ and
$c^{'}_{n}x_{n+2}+b^{'}_{n}x_{n+1}+a_{n}x_{n}=0$.

In the DRE let $y_{n}=\frac{x_{n}}{x_{n+1}}+d_{n}$. After some easy algebra
and with the crucial choice $d_{n+1}=-b_{n}/a_{n}$ one gets
$$
(c_{n}+\frac{b_{n}}{a_{n}})x_{n+2}-(1+a_{n}\frac{b_{n-1}}{a_{n-1}})x_{n+1}+
a_{n}x_{n}=0~.
\eqno(6)
$$
Thus, the identifications are $b_{n}^{'}=-1-\frac{a_{n}b_{n-1}}{a_{n-1}}$
and $c_{n}^{'}=c_{n}+\frac{b_{n}}{a_{n}}$.
Vice versa, let us start with the HSODE and introduce
$y_{n}=x_{n}/x_{n+1}+(b_{n}^{'}+1)/a_{n}$. Again, after quite simple algebra,
one gets the DRE with the following identifications
$b_{n}=-a_{n}(1+b_{n+1}^{'})/a_{n+1}$ and $c_{n}=c_{n}^{'}+
(1+b_{n+1}^{'})/a_{n+1}$. These are the inverse identifications of the
first ones.

One may foresee many physical applications of the above mathematical procedure.
Here, we present only one application,
namely to write down the general 
solution of the Hermitian three-site master equation written as in
[\refcite{j}]
and [\refcite{rr}]
$$
-(g_{n+1}r_{n+2})^{1/2}x_{n+2}+
(g_{n+1}+r_{n+1})x_{n+1}-(g_{n}r_{n+1})^{1/2}x_{n}=0~,
\eqno(7)
$$
where $g_{n}$ is the transition rate for the jump $n\rightarrow n+1$ and
$r_{n}$ is the one for the backward jump $n\rightarrow n-1$, while
$x_{n}$ can be thought of as the square root of the site probability.
The known stationary solution is [\refcite{G}]
$$
x_{n}^{0}= A
\Bigg(\prod _{j=0}^{n} \frac{g_{j-1}}{r_{j}}\Bigg)^{1/2}~,
\eqno(8)
$$
where $A$ is a scaling constant, that through the normalization
condition can be fixed to be $A=\Bigg(1+\sum _{n=1}^{N}
\prod _{j=1}^{n} \frac{g_{j-1}}{r_{j}}\Bigg)^{-1/2}~$ (see [\refcite{gil}]).
The transformation
$$
y_{n}=\frac{x_{n}}{x_{n+1}}-\frac{1+g_{n+1}+r_{n+1}}{(g_{n}r_{n+1})^{1/2}}
\eqno(9)
$$
applied to Eq. (7) leads to the following master Riccati equation
$$
y_{n+1}=-(g_{n}r_{n+1})^{1/2}y_{n+1}y_{n}-
\Bigg(\frac{g_{n}r_{n+1}}{g_{n+1}r_{n+2}}\Bigg)^{1/2}(1+g_{n+2}+r_{n+2})y_{n}
-(g_{n+1}r_{n+2})^{1/2}-\frac{1+g_{n+2}+r_{n+2}}{(g_{n+1}r_{n+2})^{1/2}}~,
\eqno(10)
$$
with the particular solution
$$
y_{n}^{0}=-\frac{1+g_{n+1}}{(g_{n}r_{n+1})^{1/2}}~.
\eqno(11)
$$
However, one can write the general solution corresponding to Eqs.~(5a,b),
where $y_{n}^{0}$ is substituted by Eq. (11) and $a_{k}=a_{k}^{M}\equiv
-(g_{k}r_{k+1})^{1/2}$
and $b_{k}=b_{k}^{M}\equiv
-\Bigg(\frac{g_{k}r_{k+1}}{g_{k+1}r_{k+2}}\Bigg) ^{1/2}
(1+g_{k+2}+r_{k+2})$, where the superscript $M$ stands for master.
By introducing the master DIF
$$
f^{M}_{p}=\prod _{l=0}^{p}\frac{b_{l}^{M}+a_{l}^{M}y_{l+1}^{0}}
{1-a_{l}^{M}y_{l}^{0}}~,
\eqno(12)
$$
one can write a compact form of the general master Riccati solution
$$
y_{n}^{1}=y_{n}^{0}+\frac{f_{n-1}^{M}}
{D-\sum _{k=0}^{n-1}\frac{f_{k}^{M}a_{k}^{M}}
{b_{k}^{M}+a_{k}^{M}y_{k+1}^{0}}}=
-\frac{1+g_{n+1}}{(g_{n}r_{n+1})^{1/2}}+
\frac{
\frac{(g_{0}r_1)^{1/2}}{g_{n}r_{n+1}}
(\frac{x_1^{0}}{x_{n+1}^{0}})^{2}}
{D-\sum _{s=0}^{n-1}
\frac{(g_0 r_1)^{1/2}}{r_{s+2}}
(\frac{x_1^0}{x_{s+2}^{0}})^{2}
}~,
\eqno(13)
$$
where $D$ is a constant.
Moreover, using simple discrete algebra, one can obtain the following
new general solution of the master equation
$$
x_{n}^{1}=x_{n}^{0}\prod _{j=0}^{n-1}
\Bigg(1+
\frac{
\frac{(g_{0}r_1)^{1/2}}{r_{j+1}}
(\frac{x_1^{0}}{x_{j+1}^{0}})^{2}}
{D-\sum _{s=0}^{j-1}
\frac{(g_0 r_1)^{1/2}}{r_{s+2}}
(\frac{x_1^0}{x_{s+2}^{0}})^{2}
}\Bigg)^{-1}~,
\eqno(14)
$$
where $x_{p}^{0}$ is the known product solution Eq.~(8)
having $p$ factors.

Let us see now the particular case $g_{n}=c_1$, $r_{n}=c_2$, where $c_1$ and
$c_2$ are two real numbers $\in (0,1)$, such that $c_1\leq c_2$, i.e.,
$q=\frac{c_1}{c_2}\leq 1$.
Then, the known stationary
master solution is $x_{n}^{0}=Aq^{\frac{n+1}{2}}$, the
particular Riccati solution is $y_{n}^{0}=-\frac{1+c_1}{(c_1c_2)^{1/2}}$,
whereas the general Riccati solution reads
$$
y_{n}^{1}=-\frac{1+c_1}{(c_1c_2)^{1/2}}+\frac{q^{-n}}{D-q^{-1/2}
\sum _{k=0}^{n-1}q^{-k}}~.
\eqno(15)
$$
At this point, we treat separately the two subcases as follows.

(i) For $c_1=c_2=\frac{1}{2}$ (free random walk),
the particular solution is $y_{n}^{0}=-3$
and the general solution is
$$
y_{n}^{1}=-3+\frac{1}{D-n}~.
\eqno(16)
$$

(ii) For $c_1\neq c_2$ (biased random walk), say,
$c_1=\frac{1}{2}(1-\epsilon)$ and
$c_2=\frac{1}{2}(1+\epsilon)$ so that $q=\frac{1-\epsilon}{1+\epsilon}$,
the particular solution is
$y_{n}^{0}=-\frac{3-\epsilon}{{(1-\epsilon ^2)}^{1/2}}$ and the general one
is the following
$$
y_{n}^{1}=-\frac{3-\epsilon}{(1-\epsilon ^2)^{1/2}} +
\frac{q^{-n}}{D-q^{-1/2}\frac{1-q^{-n}}{1-q^{-1}}}=
-\frac{3-\epsilon}{(1-\epsilon ^2)^{1/2}} +
\frac{ (\frac{1+\epsilon}{1-\epsilon})^{n} }{
D+\frac{(1-\epsilon ^2)^{1/2}}{2\epsilon}
[1-(\frac{1+\epsilon}{1-\epsilon})^{n}]
}~.
\eqno(17)
$$
For $\epsilon \rightarrow 0$ this solution goes into the above one for the
free random walk.

Passing to the master equation one can get in the two subcases the
following solutions:

(i) $x_{n}^{0}= A$  and $x_{n}^{1}=A(1-\frac{n}{D+1})$.
For the positivity of the solution one should take $D>n-1$. One can see that
for $D\approx n$, at large $n$, $x_{n}^{1}$ is nought. However, for
$D=\alpha n$, where $\alpha >1$, again at large $n$,
$x_{n}^{1}=A(1-\frac{1}{\alpha})$ is substantially different from $x_{n}^{0}$.
Finally, for $\alpha \gg 1$ one gets the particular solution.

(ii) $x_{n}^{0}=A(\frac{c_1}{c_2})^{\frac{n+1}{2}}$ and
after reabsorbing some constants one gets
$x_{n}^{1}= x_{n}^{0}\Bigg(\frac{D-\frac{1-
(\frac{c_2}{c_1})^{n+1}}{1-\frac{c_2}{c_1}}}{D-1}\Bigg)$.
Inspection of the latter formula shows that for positivity one should take
$D>\frac{(1+p)^{n+1}-1}{p}$, where $p=\frac{2\epsilon}{1-\epsilon}$, which
is close to $D>n+1$ resulting from the first order approximation.

In summary, we derived the explicit form of the
general solution of the three-site master equation based on the
corresponding discrete Riccati general solution.
In the continuous case it corresponds to
Mielnik's procedure in supersymmetric quantum mechanics.
Since the steady state
solution is always a particular solution,
what we have obtained here are either transient solutions [\refcite{M}] or
solutions determined by some nonequilibrium boundary conditions
fixing the value of the constant $D$.

\nonumsection{Acknowledgement}
\noindent
This work was supported in part by the CONACyT Project
458100-5-25844E.


\nonumsection{References}


\end{document}